\documentstyle[12pt,epsf]{article} 
\begin{document} 
\begin{titlepage} 
\begin{flushright} 
ULB--TH 98/07 
\end{flushright} 
\vspace*{1.6cm} 
 
\begin{center} 
{\Large\bf Boron abundance and solar neutrino spectrum distortion}\\ 
\vspace*{0.8cm} 
 
R.~Escribano\footnote{Chercheur IISN.}, 
J.-M.~Fr\`ere\footnote{Directeur de recherches du FNRS.}, 
A.~Gevaert and D.~Monderen\\ 
\vspace*{0.2cm} 
 
{\footnotesize\it Service de Physique Th\`eorique, Universit\'e Libre de 
Bruxelles, CP 225, B-1050 Bruxelles, Belgium} 
\end{center} 
\vspace*{1.0cm} 
 
\begin{abstract} 
The presence of neutrinos from Boron decay in the flux observed on Earth 
is attested by the observation of their energy spectrum. Possible distortions 
of the spectrum investigated in current detectors are often interpreted in 
terms of evidence in favour or against various schemes of neutrino 
oscillations. 
We stress here that a distortion of the spectrum at high energies could also
result from an increase in the ratio of neutrinos originating from ($^3$He+p) 
and $^8$B reactions. 
While a $^8$B neutrino depletion would contribute to this effect, an increase 
in the Hep contribution seems also needed to reproduce the preliminary data. 
\end{abstract} 
\end{titlepage} 
 
\section{Introduction} 
We want to study the effects of possible changes in the ratio between 
the fluxes of solar neutrinos produced respectively by the ($^3$He+p) 
reaction and by Boron decay. For simplicity, we will refer to this ratio by 
the shorthand $Hep/B$. 
 
In particular, an increase in $Hep/B$ could account for the increase in the 
number of neutrinos observed in the high-energy part of the spectrum, as 
suggested by the preliminary data of Super-Kamiokande. This is a crucial 
point to investigate, as such an increase, thus far interpreted as a 
distortion of the Boron neutrinos spectrum, is the only direct evidence (by 
this, we mean largely independent of the solar models) for solar neutrino 
oscillations. 
 
Boron abundance in the Sun has been considerably discussed since its 
energetic decay neutrinos play a leading role in most experiments, far out 
of proportion to their sheer numbers. Furthermore, the $^{8}$B reduction 
mechanism depends on the poorly known $^{7}$Be$(p,\gamma )^{8}$B production 
cross section\footnote{ 
For a recent reevaluation of this important quantity see Ref.~\cite{HAM}, 
and references therein.}. Although the close relation between Boron and 
Beryllium abundances makes it unlikely to account for all observations by a 
reduction of the $^{8}$B abundance alone \cite{HAT}, the impact of a shift 
in $^{8}$B abundance on the spectrum distortion, and the importance of the 
latter in discriminating among oscillation schemes makes it an essential 
element of a complete analysis. 
 
We began with the question: assuming that the apparent depletion of 
Boron-produced neutrinos is genuine (i.e.~not due to oscillations) would the 
corresponding change in spectra  effectively mimic the Super-Kamiokande 
signal? This is indeed largely the case, as we see in the first figure 
below, if the comparison is made directly between the inferred electron 
recoil curves and the preliminary data. It turns out, however, that a severe 
smoothing occurs, due to the limited energy resolution of the experiment, 
and this must be included in the comparison. This is done in the second 
figure, which shows clearly that a much larger increase in $Hep/B$ is needed 
to reproduce the data. Such a large increase cannot stem from a reduction in 
the Boron contribution alone, as such a suppression would contradict the 
data. Instead the possibility of an enhancement of the Hep contribution, 
either for nuclear or astrophysical reasons, must be called into play. 
 
Even apart from possible astrophysical effects, it turns out indeed  that 
the ($^3$He+p) reaction is in fact poorly known, and could strongly 
increase $Hep/B$. 
 
After discussing the effect of varying $Hep/B$, we take the opportunity to 
review in simple terms how it would interfere with the expected spectrum 
distortions in various oscillation schemes 
 
\section{Varying $^{8}$B and Hep neutrino abundances} 
The energy range explored by the Super-Kamiokande experiment is dominated by 
neutrinos from $^{8}$B decay \cite{SKAK}. In the upper part of the spectrum, 
however, this spectrum crosses the contribution from Hep neutrinos\footnote{ 
Hep neutrinos are those produced in the nuclear reaction  
$^3$He$+p\rightarrow ^4$He$+e^++\nu_e$.}. Clearly, despite a higher energy 
and thus better sensitivity, this upper part is more difficult to measure as 
the absolute flux drops by orders of magnitude, so that the statistical 
significance dwindles. 
 
\begin{figure} 
$$ 
\epsffile{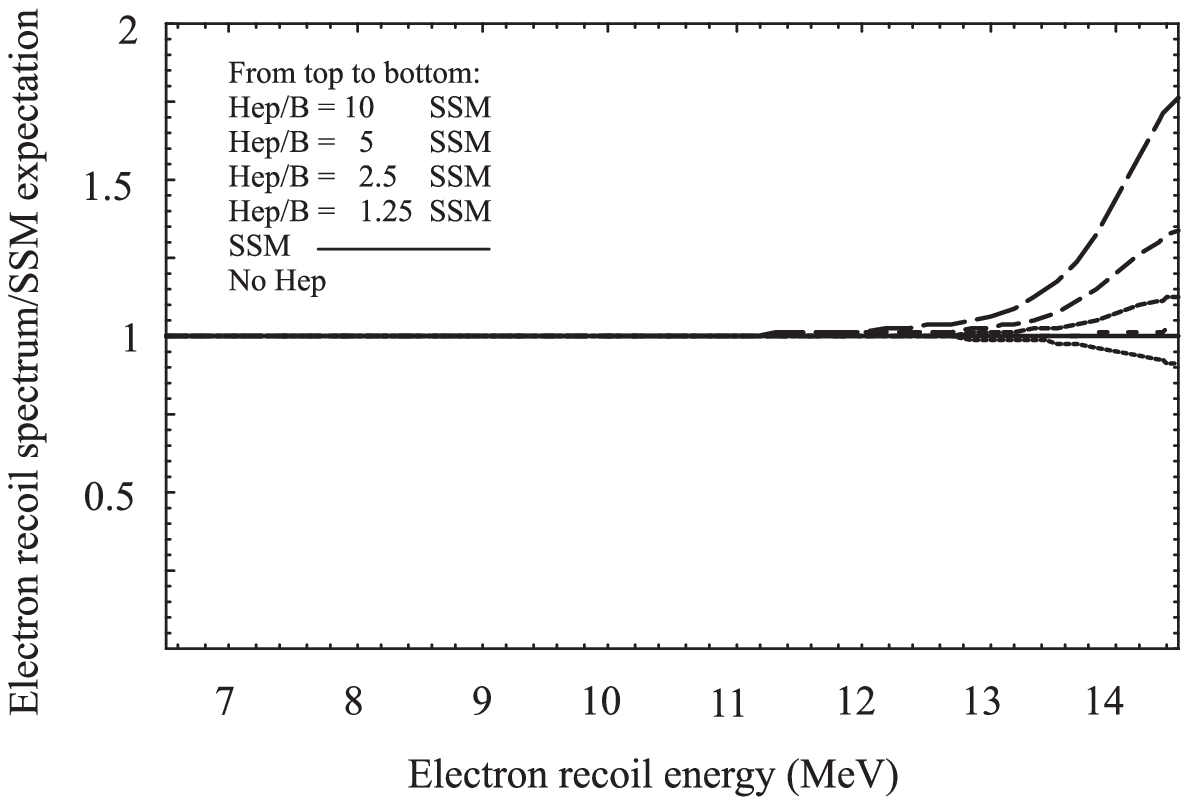} 
$$ 
\caption{Electron recoil spectra as a function of $^8$B abundance in the 
Sun. The solid line refers to the SSM, to which all curves are normalised.} 
\label{fig1} 
\end{figure}

We have plotted in Fig.~\ref{fig1} the expected electron recoil spectrum 
versus the energy for various $Hep/B$ ratios. The plot is based on SSM data 
and standard differential cross sections \cite{BAH}. 
 
The curves are normalized to give an equal number of events above the 
Super-Kamiokande threshold, and the plotted points have then been reduced to 
the SSM expectation. For completeness, we have also plotted a curve omitting 
completely the Hep contribution. In these conditions, decreasing the  
$^{8}$B abundance amounts to increasing the relative role of the Hep 
contribution, and thus in an enhancement of the expectation for large recoil 
energies. While such graphs are now standard, it may at first sight seem 
surprising that, when all the sets of data have been normalized to the same 
total number of events, increasing $Hep/B$ leads to an increase at high 
energy values, apparently not compensated for by a decrease at low 
energies. The reason for this somewhat misleading effect is simply 
understood when referring to the raw values (not normalized to SSM 
expectations). Indeed the fall in energy distribution is so steep that the 
high-energy events (and in particular the Hep contribution) have only 
a minute impact on the normalization. When the data are normalized to the 
SSM model, the depletion of the low-energy part reaches the graphical 
resolution limit. 
 
The top four curves shown in Fig.~\ref{fig1} correspond respectively to 
reductions of the $^{8}$B abundance by 0.1, 0.2, 0.4, and 0.8 of its 
standard value with respect to the Hep contribution, or alternatively to   
$Hep/B$ = 10, 5, 2.5 and 1.25.  

\begin{figure} 
$$ 
\epsffile{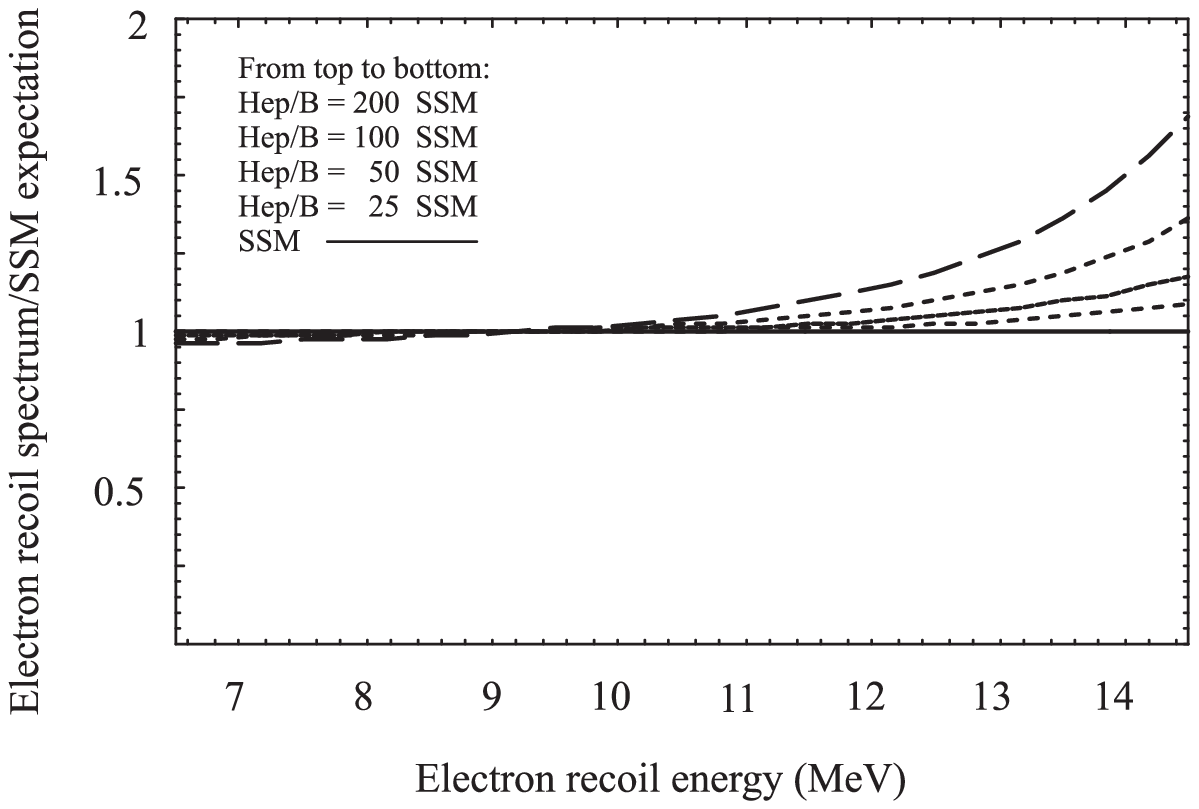} 
$$ 
\caption{Electron recoil spectra as a function of $^8$B abundance in the 
Sun, including energy resolution smoothing. The solid line refers to the 
SSM, to which all curves are normalised.} 
\label{fig2} 
\end{figure}

The curves shown in Fig.~\ref{fig1}, while close in aspect to the 
preliminary data for the larger values of $Hep/B$, are however not the ones 
effectively observed. Even discounting the habit of grouping all points 
above 14 MeV in a single bin, care has to be taken to include the effective 
energy resolution, which is usually not deconvolutioned from experimental 
presentations. This is usually done using a convolution with a gaussian-like 
resolution function \cite{BAH2}. We illustrate the effect in Fig.~\ref{fig2}.  
Obviously, the effect of Hep neutrinos is dwarfed in with respect to  
Fig.~\ref{fig1} for a given $Hep/B$, since a number of lower energy $^{8}$B 
neutrinos now inevitably ``bleed'' into the higher energy region. 

We will refrain here from reproducing the experimental points (which can 
only be gathered from figures presented at conferences, see Ref.~\cite{SKAK})  
since they have not been actually published. Suffice it for now to say 
that the higher curves in Fig.~\ref{fig2} match closely the experimental 
results, as we know them today. We also performed  tentative $\chi^{2}$ 
fits based on these preliminary data (unfortunately the error bars could 
only be taken from the figures). This indeed favours high $Hep/B$ ratios, 
giving notably a numerically better fit than the small angle MSW solution 
when fitting the spectrum in search for neutrino oscillations. Our central  
value falls between the top two curves in Fig.~\ref{fig2}. At this 
moment, however, the data accuracy is not sufficient to conclude. 
 
It is thus clearly important to allow for a (large) variation of the ratio 
of Hep to $^{8}$B contributions in the study of oscillations. 
 
While the observed neutrino flux does seem to indicate a suppression of the  
$^{8}$B with respect to SSM expectations (assuming for the moment the absence 
of oscillations) by a factor 2 to 3 \cite{SKAK2}, this is totally 
insufficient to account for the large enhancement of $Hep/B$ needed 
according to Fig.~\ref{fig2}. However, even not taking into account 
possible astrophysical effects, the ratio of the Hep to $^{8}$B neutrinos is 
further affected by the severe uncertainty on the  
$^{3}$He$(p,e^{+}\nu_{e})^{4}$He cross section \cite{ADE}.  
The central values quoted in the 
recent literature for the nuclear factor $S_{13}(0)$ of this cross section 
vary from (1.3 to 57)$\times 10^{-20}$ keV \cite{ADE}, with the lowest value 
used by \cite{BAH}, which we have adopted for normalization. 
Increase in $Hep/B$ of up to 150 with respect to \cite{BAH}, and 
resulting from both $^{8}$B depletion and Hep enhancement could thus be 
considered (even outside possible astrophysical effects).  
 
\section{Oscillation schemes} 
Oscillation effects are known to affect the shape of the electron recoil 
spectrum. We list briefly the expected effects in order to see how they 
conspire or interfere with a possible increase in the $Hep/B$ ratio. 
 
In the case of MSW oscillations, the situation differs according to the 
adiabatic or otherwise evolution of the system . In the adiabatic case, for 
a large enough value of $\frac{E}{\delta m^{2}}$ the electronic neutrinos 
are the ``heavy'' solution inside the Sun and move continuously into the 
``heavy'' solution (a $\nu _{\mu }$, $\nu _{\tau }$ or a sterile neutrino  
$\nu _{s}$) outside the Sun. Since the resonance condition is met at lesser 
densities for higher energies, the cross-over occurs at a larger solar 
radius, and thus a larger proportion of the energetic neutrinos are 
effectively (totally or partially) transformed/sterilized. The distortion is 
then expected to be negative, but could be partially hidden by the above 
abundance considerations. 
 
In the non-adiabatic case, however, the slope of the solar density close to 
the resonance point changes fast enough to allow the neutrinos to jump from 
the ``heavy'' solution to the ``light'' solution of the level-crossing 
diagram. This occurs increasingly with energy, so for small mixing at least, 
a positive slope is expected. 
 
Such a situation is usually invoked because it allows for the suppression of 
a ``slice'' of the neutrino spectrum (typically the $^{7}$Be lines) while 
restoring a sufficient  proportion of $^{8}$B neutrinos for the 
Super-Kamiokande observations. In this case, clearly, the two effects, 
i.e.~non-adiabatic oscillations and $Hep/B$ enhancement, would go in the 
same direction, but the detailed pattern is quite different. $^{8}$B 
depletion typically affects the end part of the spectrum, while the effect 
of non-adiabaticity is rather gradual. Higher statistics will be needed to 
disentangle the two effects. 
 
Finally, a possible solution in the shape of long-range (mostly vacuum) 
oscillations has been discussed recently \cite{SMI}. Here, the distortion of 
the spectrum depends on the position of the oscillation curve  
$\sin^{2}\left( \frac{\delta m^{2}L}{4E}\right)$, that is, on the number of 
oscillations between the Sun and the Earth. Both positive and negative 
slopes are then possible. As pointed out in \cite{SMI}, the two effects will 
be distinguishable through different correlations with the Sun-Earth 
distance, which only affects the oscillation part. 
 
\section{Conclusions} 
Future experiments and extra statistics from Super-Kamiokande will improve 
understanding of the solar neutrino problem. In comparing with models, we 
insist, however, that both the ratio $Hep/B$ of Hep to $^{8}$B neutrinos (as 
well as its implications for other experiments) and various oscillation 
schemes must be considered simultaneously. Preliminary values favour a large 
ratio of Hep to $^{8}$B neutrinos. Alternatively, a cut in the energy 
spectrum might allow a nearly independent study of both effects, since  
$^{8}$B depletion and/or Hep enhancement show up significantly only in the 
higher parts of the energy spectrum. 
 
\section{Acknowledgements} 
This work was supported by I.~I.~S.~N.~Belgium. D.~Monderen benefits from a 
FRIA grant. We thank P.~Vilain and G.~Wilquet for numerous discussions.

\end{document}